\documentclass{article}
\usepackage{spconf,amsmath,graphicx}
\usepackage[utf8]{inputenc}
\usepackage{import}
\usepackage{color}
\usepackage{pstricks} 
\usepackage{tabularx, multirow}

\usepackage{calc}  
\usepackage{enumitem}

\usepackage{siunitx}

\ninept

\title{Iterative optimization of quarter sampling masks for Non-Regular Sampling Sensors}
\name{Simon Grosche, Jürgen Seiler and André Kaup}
\address{Chair of Multimedia Communications and Signal Processing\\
University of Erlangen-Nuremberg, Cauerstr. 7, 91058 Erlangen, Germany\\
 \textit{\small \{simon.grosche, juergen.seiler, andre.kaup\}@fau.de }
}

\usepackage[firstpage=true]{background}
\usepackage{hyperref}

\SetBgContents{\parbox{\textwidth}{\footnotesize © 2018 IEEE. Personal use of this material is permitted. Permission from IEEE must be
		obtained for all other uses, in any current or future media, including
		reprinting/republishing this material for advertising or promotional purposes, creating new
		collective works, for resale or redistribution to servers or lists, or reuse of any copyrighted
		component of this work in other works. DOI: \href{https://doi.org/10.1109/ICIP.2018.8451658}{10.1109/ICIP.2018.8451658.}}}
\SetBgScale{1}
\SetBgAngle{0}
\SetBgPosition{current page.south}
\SetBgVshift{1cm}
\SetBgColor{black}
\SetBgOpacity{1}

\begin{document}
\maketitle
\begin{abstract}
Non-regular sampling can reduce aliasing at the expense of noise. Recently, it has been shown that non-regular sampling can be carried out using a conventional regular imaging sensor when the surface of its individual pixels is partially covered. This technique is called quarter sampling (also 1/4 sampling), since only one quarter of each pixel is sensitive to light.
For this purpose, the choice of a proper sampling mask is crucial to achieve a high reconstruction quality.
In the scope of this work, we present an iterative algorithm to improve an arbitrary quarter sampling mask which results in a continuous increase of the reconstruction quality.
In terms of the reconstruction algorithms, we test two simple algorithms, namely, linear interpolation and nearest neighbor interpolation, as well as two more sophisticated algorithms, namely, steering kernel regression and frequency selective extrapolation.
Besides PSNR gains of \SIrange[retain-explicit-plus]{+0.31}{+0.68}{\decibel} relative to a random quarter sampling mask resulting from our optimized mask, visually noticeable enhancements are perceptible.
\end{abstract}
\section{Introduction}
\vspace*{-0.1cm}
\label{sec:intro}

Non-regular sampling of images can be used to reduce aliasing conventionally occurring from regular sampling \cite{Dippe1985, Hennenfent2007, Maeda2009}. Additionally, it has been suggested to use non-regular sampling to enhance the spatial resolution of an imaging sensor~\cite{Schoberl2011}. This so called quarter sampling (1/4 sampling) can increase the resolution of an image by physically masking three quarters of each pixel of a low resolution sensor. While the same amount of energy, cost and data bandwidth is needed as for the low resolution sensor, the higher resolution is achieved at the expense of an additional post-processing step performing the extrapolation on the high resolution grid which features twice the resolution in both dimensions.

\begin{figure}[t]
	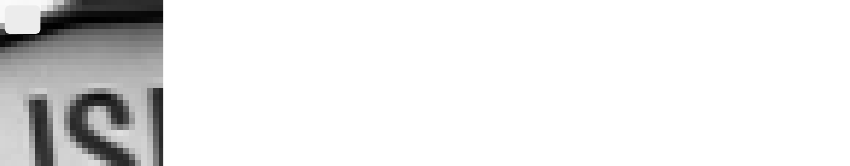
	\vspace*{-7mm}
	\caption{(a) Detail view of a high resolution image. (b) Simulated low resolution image by averaging $2{\times} 2$ pixels from (a), respectively. (c) Random quarter sampling mask. The horizontal and vertical red lines indicate the borders of the low resolution pixels. The white areas are transparent. (d) Simulated measurement with the masked sensor.
	(e) Reconstructed high resolution image. Here the FSR is used. \textit{(Please pay attention, additional aliasing may be caused by printing or scaling. Best to be viewed enlarged on a monitor.)}}
	\label{fig:sampling_detailedview}
	\vspace*{-2mm}
\end{figure}

Figure\,\ref{fig:sampling_detailedview} depicts the concept behind quarter sampling:  Fig.\,\ref{fig:sampling_detailedview}(a) shows a high resolution test image taken with a high resolution sensor with twice the resolution of our presumed low resolution sensor. In a simplified model, the low resolution sensor measures a downscaled version of that image by averaging $2{\times} 2$ pixels at a time. A simulated measurement of the low resolution sensor is shown in Fig.\,\ref{fig:sampling_detailedview}(b). Conversely, when quarter sampling is used three quarters of each pixel of the low resolution sensor are covered.
Figure\,\ref{fig:sampling_detailedview}(c) shows a random quarter sampling mask. Each of these pixels is covered by three quarters (black) and only one quarter of the pixel surface is transparent such that the measurement can be describes as a sub-sampling of the image in Fig.\,\ref{fig:sampling_detailedview}(a). The sub-sampled image is depicted in Fig.\,\ref{fig:sampling_detailedview}(d). Finally, the missing pixels on the high resolution grid need to be reconstructed yielding an image as in Fig.\,\ref{fig:sampling_detailedview}(e). It is important to note that the complete high resolution data is not present in an actual hardware implementation of the sensor because only the data depicted in Fig.\,\ref{fig:sampling_detailedview}(d) is measured and used for the reconstruction.
Therefore, Fig.\,\ref{fig:sampling_detailedview}(b) and Fig.\,\ref{fig:sampling_detailedview}(e) have the same number of sampling data points but placing them non-regularly leads to higher image quality.

It turns out that the resulting image quality depends on the chosen quarter sampling mask.
For example, a regular quarter sampling mask, which contains the transparent area in the same corner for each low resolution pixel, is disadvantageous since it leads to aliasing again. On the other hand, random quarter sampling masks are expected to be non-optimal since they may contain large covered areas which are in turn harder to reconstruct.

In terms of the reconstruction algorithm we investigate linear interpolation and nearest neighbor interpolation, as well as two more sophisticated algorithms, namely, steering kernel regression (SKR)~\cite{Takeda2007} and frequency selective reconstruction (FSR)~\cite{Seiler2015}. FSR has shown to be a successful reconstruction scheme for various inpainting and extrapolation tasks \cite{Herraiz2008, Stehle2006,Seiler2009} and showed best results for non-regular sampling and quarter sampling in~\cite{Schoberl2011,Seiler2015}.

This work is organized as follows: Section\,\ref{sec:existing_opt} summarizes existing optimization strategies for quarter sampling masks. In Section\,\ref{sec:mask_opt}, we identify properties of a good quarter sampling mask and use these properties to propose an algorithm which iteratively improves the quality of an arbitrary quarter sampling mask. This leads to high quality masks of size $8{\times} 8$ and $32{\times} 32$. In Section\,\ref{sec:eval_masks}, the reconstruction quality using these masks is evaluated. Therein, visual comparisons are presented as well.

\section{Existing mask optimization strategies}
\vspace*{-0.1cm}
\label{sec:existing_opt}

In order to find an optimal quarter sampling mask, Jonscher\,et.al. suggest to use a brute force method~\cite{Jonscher2014}.
They suggest to randomly generate quarter sampling masks of size $b{\times} b$, $b \in \left\{2,4,8,16,\dots\right\}$ and repeat them periodically to match the image size on the high resolution grid.
The masks are then applied to a set of test images.  Afterwards,  high resolution images are reconstructed from the sub-sampled data and the PSNR can be calculated.
The mask yielding to the highest average PSNR for a given reconstruction algorithm is chosen for future sampling tasks. Of course, care has to be taken when choosing the test images as these need to be representative.

Unfortunately, there is a serious downside with this approach. For each of the $(b/2)^2$ low resolution pixels on a quarter sampling masks of size $b{\times} b$, there are four possible choice to place the transparent pixel. Therefore, the number of possible masks is
$N_b{=}4^{b^2 / 4}$
and thus scales exponentially with the number of pixels. The actual values are $N_2 \,{=}\, 4, N_4 \,{=}\, 256, N_8\,{\approx}\,\num{4e9}, N_{16}\,{\approx}\,\num{3e38},  ... \,$. For large masks ($b\ge 8$), the number of all possible masks makes it computationally infeasible to generate all possible masks and perform the reconstruction on several sub-sampled test images.

Jonscher\,et.al.\ incorporate this issue by choosing 256 randomly selected quarter sampling masks for each $b\ge4$. This means, that only a small subset of all possible masks is tested for $b\ge 8$.
Then, they evaluate all chosen masks on a set of test images and find the mask with the best average PSNR for each $b$ and each investigated reconstruction algorithm. For example, for the FSR their best mask found is of size $b=8$.
Even though their brute force method theoretically works, it is practically unsatisfactory, because only a small subset of masks is tested for $b\ge 8$.
Since all masks of size $b$ are included in the set of all masks of size $b'=2\cdot b$. the reconstruction quality cannot decrease for larger masks, if all possible masks are taken into account.

To overcome the problems with this brute force method, we propose an optimization strategy in the next section.
We circumvent calculating all possible masks and additionally do not need to perform any reconstruction during the mask optimization. It is not our goal to find the one and only best mask for a given reconstruction algorithm but rather to provide a method to find a very good mask based on reasonable heuristic assumptions.

\section{Proposed mask optimization}
\vspace*{-0.1cm}
\label{sec:mask_opt}
\subsection{Properties of a good mask}
\vspace*{-0.05cm}
Before being able to optimize a quarter sampling mask, we need to identify properties to distinguish favorable and unfavorable masks.

\begin{figure}[t]
	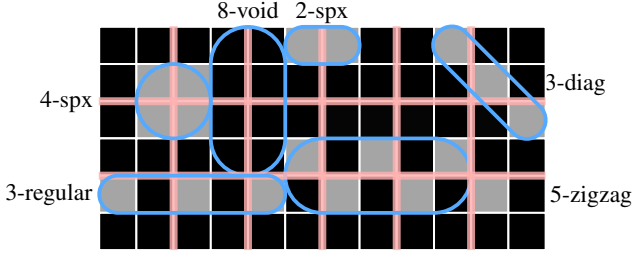
	\vspace*{-7mm}
	\caption{Quarter sampling mask of size $6{\times}12$ with the structures we identified as unfavorable being marked by blue rounded rectangles. The horizontal and vertical red lines indicate the borders of the low resolution pixels. Black areas are masked, whereas the gray areas are transparent.}
	\label{fig:cluster_being_removed}
	\vspace*{-4.5mm}
\end{figure}

We claim that two properties are favorable for a good quarter sampling mask: (A) low regularity and (B) uniformity.
(A) Regularity leads to high peaks in the spectrum of the masking function. As an immediate consequence, aliasing occurs. This is unfavorable since information about the image is irretrievable lost.
In order to achieve a low regularity, a random quarter sampling mask could be chosen.
(B) Uniformity is the second desired property. Within an image, details consisting of very few pixels can be anywhere with the same probability and should be captured independent of their location.
This condition implies that the mask should be as uniform as possible. While a regular mask means high uniformity, a random mask means low uniformity.
Putting these two properties together, both the random mask and the regular masks are extreme cases complying only with one of the properties.

\vspace*{-0.075cm}
\subsection{Proposed optimization strategy}
\vspace*{-0.05cm}
\label{sec:mask_opt_strat}

An optimized mask should combine both properties, i.e., it should be both non-regular and uniform.
To achieve this, we propose that an arbitrary quarter sampling mask can be improved by reducing the occurrences of the following structures (see also Fig.\,\ref{fig:cluster_being_removed}):
\vspace{-0.4\topsep}
\begin{description}[leftmargin=0cm, labelwidth=0cm]
	\setlength{\itemsep}{0pt}
	\setlength{\parskip}{0pt}
	\setlength{\parsep}{0pt}
	\item[2-spx] horizontal/vertical pair of transparent pixels forming a  superpixel (spx).
	\item[4-spx] pair of two 2-spx forming a  transparent $2{\times}2$ superpixel, already mentioned in \cite{Jonscher2014}.
	\item[8-void] $2{\times}4$ or $4{\times}2$   block of masked pixels forming a large unknown area.
	\item[3-regular] three transparent pixels in a horizontal/vertical line, spaced with two masked pixels.
	\item[3-diag] three neighboring pixels lying on a diagonal.
	\item[5-zigzag] five pixels in a horizontal/vertical zigzag assembly.
\end{description}
\vspace{-0.4\topsep}

Removing the superpixel structures (2\hbox{-}spx, 4\hbox{-}spx) and the void structure (8\hbox{-}void) results in a more uniform mask, while the removal of the regular structures (3\hbox{-}regular, 3\hbox{-}diag and 5\hbox{-}zigzag) makes the mask less regular.

\begin{figure}[t]
	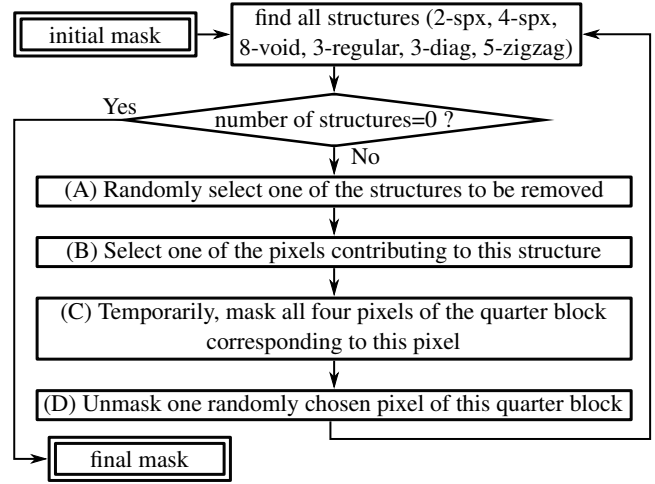
	\vspace*{-7mm}
	\caption{Flow graph of the proposed mask optimization algorithm.}
	\label{fig:flow_graph}
	\vspace*{-5mm}
\end{figure}

\begin{figure}[t]
	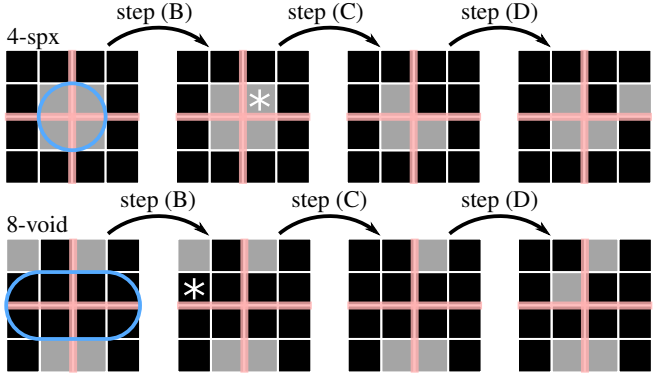
	\vspace*{-7mm}
	\caption{Example for a 4\hbox{-}spx and a 8\hbox{-}void structure being removed. For detailed explanation of the algorithm, see text.}
	\label{fig:struc_remove}
	\vspace*{-2mm}
\end{figure}

To produce masks with a reduced number of these structures from an arbitrary quarter sampling mask, we propose an iterative algorithm consisting of four core steps (A-D) to remove the structures. A flow graph illustrating the whole algorithm is depicted in Fig.\,\ref{fig:flow_graph}.
The steps are  exemplarily shown for the removal of a 4\hbox{-}spx and a 8\hbox{-}void structure in Fig.\,\ref{fig:struc_remove} and can be described as follows:
\vspace{-0.4\topsep}
\begin{itemize}[leftmargin=!, labelwidth=\widthof{(D)}]
\setlength{\itemsep}{0pt}
\setlength{\parskip}{0pt}
\setlength{\parsep}{0pt}
	\item[(A)] First, one randomly selected structure of those depicted in Fig.\,\ref{fig:cluster_being_removed} is selected to be removed.
	\item[(B)] Then, one of the pixels contributing to this structure is randomly selected (marked with an asterisk in Fig.\,\ref{fig:struc_remove}).
	\item[(C)] Temporarily, all pixels of its corresponding quarter-block are masked.
	\item[(D)] Finally, a randomly selected pixel of that quarter-block is unmasked, such that the quarter sampling condition is fulfilled.
\end{itemize}
\vspace{-0.4\topsep}
The steps (A)-(D) are repeated over and over again. With this algorithm, the structures are removed most of the time while it actually allows to reproduce them. Furthermore, the algorithm potentially creates structures of a different type that were not present before. Nevertheless, the total number of structures is expected to decrease in average.
\begin{figure}[t]
	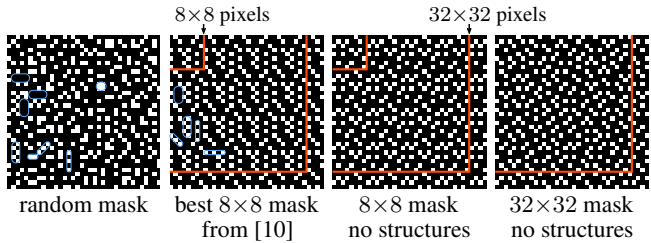
	\vspace*{-8mm}
	\caption{Different quarter sampling masks (repeated periodically if necessary) cropped to $36{{\times}} 36$ pixels. Whereas the random mask features many structures (some marked with blue rounded rectangles) and does not have a uniform appearance, our optimized masks are structure-free and have a uniform appearance while still being non-regular.}
	\label{fig:masks_random_Jonscher_nocluster}
	\vspace*{-5mm}
\end{figure}
During the search and the removal of the structures periodic boundary conditions have to be taken into account, since masks being smaller than the size of the image are repeated periodically.

Now, our goal is to generate a quarter sampling mask without any structures to reach a high quality of uniformity and non-regularity. 
In terms of the desired size of the quarter sampling masks, we will restrict ourselves to sizes up to $b{\times} b=32{\times}32$  high resolution pixels. This simplification can be made because all investigated reconstruction algorithms act locally, i.e., the masked data is reconstructed from its immediate vicinity. However, our algorithm is capable of creating larger  masks, if necessary. As a side effect, the periodicity of the masks may allow for a simpler manufacturing process of the actual quarter sampling sensor.

To generate a quarter sampling mask without any structures, a $32{\times} 32$ random quarter sampling mask is used as initial mask and optimized using the just described algorithm. In the beginning, the number of remaining structures goes down quickly because removing one structure might even remove another structure simultaneously. Conversely, removing the last remaining structures is more difficult, because with every removal new structures may be generated. Later, these new structures need to be removed in addition to the remaining ones. The optimization is stopped, after all structures have been removed. Finally, a $32{\times} 32$ quarter sampling mask without any structures is found.
Similarly, we  created a additional quarter sampling mask of size $8{\times} 8$ without any structures.

Figure\,\ref{fig:masks_random_Jonscher_nocluster} shows a random quarter sampling mask, the best $8{\times} 8$ mask from~\cite{Jonscher2014} and our two just described masks. While the random mask contains all types of structures, the best $8{\times} 8$ mask from~\cite{Jonscher2014} contains only 8\hbox{-}void, 3\hbox{-}diag, 3\hbox{-}regular and 5\hbox{-}zigzag and our $8{\times} 8$ mask as well as our $32{\times} 32$ mask contains no structures at all.

\vspace*{-0.075cm}
\section{Evaluation of the masks}
\vspace*{-0.07cm}
\label{sec:eval_masks}
\subsection{Validation of the iterative algorithm}
\vspace*{-0.075cm}

\begin{figure}[t]
	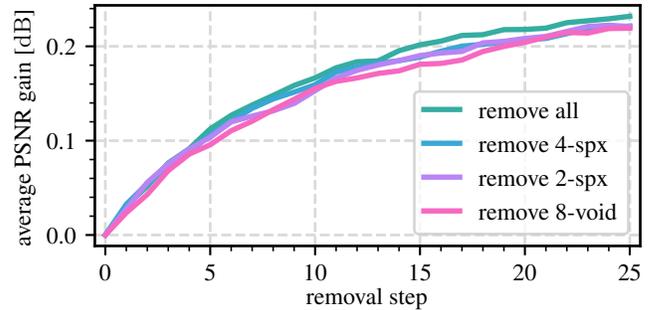
	\vspace*{-8mm}
	\caption{PSNR (averaged over the first ten TECNICK images using FSR for the reconstruction) as a function of stepwise reduction of different structures. The PSNR increases steadily as more and more structures are removed -- as expected. Furthermore, removing all structures within each step leads to the best results.}
	\label{fig:psnr_vs_itermask}
	\vspace*{-5mm}
\end{figure}

In order to test the performance of our algorithm, we perform an experiment with a random quarter sampling mask and stepwise remove more and more structures. After each step, the PSNR averaged over the first ten TECNICK images~\cite{Asuni2014} using the FSR as an exemplarily  reconstruction algorithm is calculated. 
Figure\,\ref{fig:psnr_vs_itermask} shows the average PSNR for 26 different masks of size $256{\times} 256$ where in each step 5-10\,\% of the structures were randomly selected and removed as explained in Section\,\ref{sec:mask_opt_strat}. To separate the contributions of the different structures, only the 8\hbox{-}voids, only the 2\hbox{-}spx, only the 4\hbox{-}spx and all structures were removed, respectively. All cases show that the average PSNR increases more than \SI{0.2}{\decibel} after 25 steps. When only one of the other structures (3\hbox{-}regular, 3-diag and 5\hbox{-}zigzag) is removed, the PSNR does not increase. This is due to the fact that when removing those regular structures, too many other structures like 8\hbox{-}void, 4\hbox{-}spx and 2\hbox{-}spx are generated. Conversely, when only removing 8\hbox{-}void, 4\hbox{-}spx and 2\hbox{-}spx a high number of 3\hbox{-}regular, 3-diag and 5\hbox{-}zigzag structures are generated, which finally leads to a locally regular mask. This is again disadvantageous and limits the maximally achievable reconstruction quality.
For this reason, it is most effective and leads to the highest average PSNR in our experiment, when a fraction of all structures is removed in every step.

\vspace*{-0.075cm}
\subsection{Reconstruction quality using our structure-free masks}
\vspace*{-0.075cm}
In order to evaluate the quality of our structure-free quarter sampling masks, we compare our masks with the best $8{\times} 8$ mask using the FSR from~\cite{Jonscher2014} and a random mask (all depicted in Fig.\,\ref{fig:masks_random_Jonscher_nocluster}). For completeness, the results are also provided for a regular mask. We use linear interpolation, nearest neighbor interpolation (both from \cite{scipy}), steering kernel regression (SKR)~\cite{Takeda2007} and frequency selective reconstruction (FSR) \cite{Seiler2015} for the reconstruction of the unknown pixels on the high resolution grid. All PSNR values are averaged over all 100 images of the TECNICK image database~\cite{Asuni2014}.
Since these images are taken with a regular sensor, we simulate a quarter sampling sensor by sub-sampling the images according to the quarter sampling masks (as in Fig.\,\ref{fig:sampling_detailedview}). The reconstructed images can finally be compared with the original image.

\begin{figure}[t]
	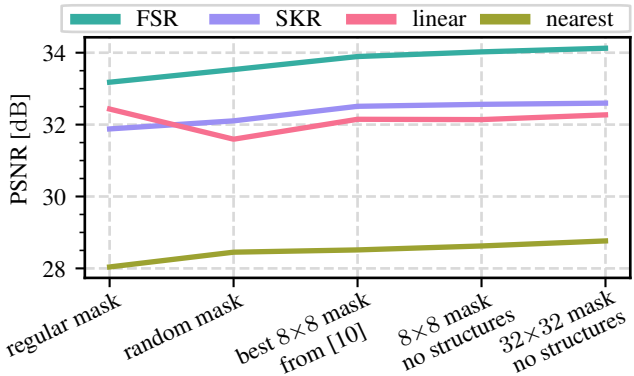
	\vspace*{-7mm}
	\caption{PSNR in \si{\decibel} (averaged over all 100 TECNICK images) for different masks and reconstruction algorithms.}
	\label{fig:psnr_vs_mask_random_8x8Jonscher_8x8_32x32_absolut}
	\vspace*{-4mm}
\end{figure}

Table\,\ref{tab:psnr_vs_mask_random_8x8Jonscher_8x8_32x32} and Fig.\,\ref{fig:psnr_vs_mask_random_8x8Jonscher_8x8_32x32_absolut} present the PSNR for the four masks depicted in Fig.\,\ref{fig:masks_random_Jonscher_nocluster} and a regular mask.
For all reconstruction algorithms we tested, the PSNR increases from the random quarter sampling mask via the best $8{\times} 8$ mask from~\cite{Jonscher2014} to our $32{\times} 32$ mask without structures. The PSNR gains are between \SI[retain-explicit-plus]{+0.31}{\decibel} and \SI[retain-explicit-plus]{+0.68}{\decibel}. For linear interpolation also the regular mask shows an improvement, whereas in case of the other reconstruction algorithms the regular mask is an unfavorable choice. In case of FSR, this can be explained with the regularity yielding to high contributions in the spectrum of the masked image which finally results in disadvantageous basis functions being selected within the FSR.
On the other hand, linear interpolation seems to benefit most from the uniformity of the regular mask compared to other masks. However, for images with high frequency contributions, the regular mask will create severe aliasing and the high resolution data cannot be reconstructed with any of the algorithms as the data is irretrievable lost due to aliasing.
\begin{table}[t]
	\vspace*{-1.8mm}
	\centering
	\setlength{\tabcolsep}{0.3pt}
	\caption{PSNR in \si{\decibel} (averaged over all 100 TECNICK images) for a regular mask, a random mask, the best $8{\times} 8$ mask from~\cite{Jonscher2014}, our $8{\times} 8$ mask without structures and our $32{\times} 32$ mask without structures. Values in parentheses provide the PSNR gains relative to the random mask. Bold font indicates the highest PSNR in each line.}
	\label{tab:psnr_vs_mask_random_8x8Jonscher_8x8_32x32}
	\vspace*{0.1mm}
	\begin{tabularx}{\columnwidth}{rccccc}
		& \rotatebox{0}{\parbox{1cm}{\centering \phantom{phant} regular\\ mask}} & \rotatebox{0}{\parbox{1cm}{\centering \phantom{phant} random\\ mask}} & \rotatebox{0}{\parbox{2cm}{\centering \phantom{phant} best $8{\times} 8$\\ mask in\,\cite{Jonscher2014}}} & \rotatebox{0}{\parbox{1.7cm}{\centering our\\$8{\times} 8$ mask\\ no structures }} & \rotatebox{0}{\parbox{2cm}{\centering our\\$32{\times} 32$ mask\\ no structures}} \vspace*{0.4mm} \\ \hline\hline\rule{0pt}{2.5ex}  
		FSR & 33.18  & 33.53  & 33.89 (+0.36) & 34.02  & \textbf{34.12} (+0.59) \\ 
		SKR & 31.88  & 32.10  & 32.51 (+0.40) & 32.56  & \textbf{32.60} (+0.49)\\ 
		linear & \textbf{32.43}  & 31.59  & 32.15 (+0.55) & 32.14  & 32.27 (+0.68)\\ 
		nearest & 28.04  & 28.45  & 28.51 (+0.06) & 28.62  & \textbf{28.76} (+0.31)
	\end{tabularx}
\vspace*{-2mm}
\end{table}

Putting these evaluations together, we can summarize that our approach to find a better quarter sampling mask worked and results in an improvement of \SIrange[retain-explicit-plus]{+0.31}{+0.68}{\decibel} compared to the random quarter sampling mask. The highest PSNR is reached using the FSR independent of the chosen mask. For the FSR, using the $8{\times} 8$ ($32{\times} 32$) mask without structures yielded to an PSNR gain of \SI[retain-explicit-plus]{+0.49}{\decibel} (\SI[retain-explicit-plus]{+0.59}{\decibel}).
To put this into perspective, the best $8{\times} 8$ mask from~\cite{Jonscher2014} resulted in an  \SI[retain-explicit-plus]{+0.36}{\decibel} increase compared to the random mask using the FSR.
It should be highlighted again that the tendency of our achieved quality improvement is independent of the tested reconstruction algorithms.

\vspace*{-0.05cm}
\subsection{Visual comparison}
\begin{figure}[t]
	\vspace*{1mm}
	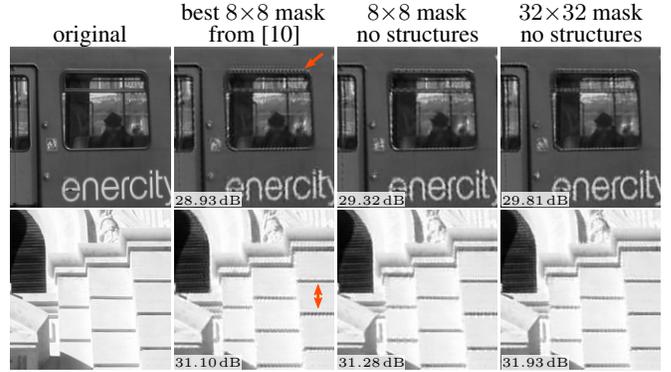
	\vspace*{-7mm}
	\caption{Visual comparison of the reconstruction quality using the FSR and different masks. Red arrows indicate regions of low quality due to non-uniformity of the best $8{\times} 8$ mask from~\cite{Jonscher2014}. The PSNR is provided for the respective image details. \textit{(Please pay attention, additional aliasing may be caused by printing or scaling. Best to be viewed enlarged on a monitor.)}}
	\label{fig:visualqualityimprovement}
	\vspace*{-4mm}
\end{figure}

\vspace*{-0.05cm}
In addition to the improvement in the PSNR, we also observe visually noticeable enhancements.
Figure\,\ref{fig:visualqualityimprovement} shows two image details reconstructed with the FSR using different quarter sampling masks. In both cases, it can be observed that some edges (marked with a small arrow) show severe artifacts with the best $8{\times} 8$ mask from~\cite{Jonscher2014}. These artifacts can be explained by the sampling mask which coincidentally consists of large voids in the region of these edges making the reconstructions less accurate. This also explains why a similar edge occurring multiple times in the exterior wall (bottom row) is sometimes reconstructed poorly and sometimes successfully. Using our $8{\times} 8$ mask without structures already reduces these artifacts. Using our $32{\times} 32$ mask without structures, no such artifacts are observed and multiple occurring edges are reconstructed with comparable quality.

\vspace*{-0.07cm}
\section{Conclusion}
\vspace*{-0.05cm}
An iterative algorithm to improve quarter sampling masks was presented within this paper. Our proposal is to remove certain structures such that the quarter sampling mask becomes less regular and more uniform.
It has been shown, that when the 8\hbox{-}void, the 2\hbox{-}spx, the 4\hbox{-}spx or all structures are stepwise removed with our approach, the reconstruction quality also increases stepwise. Furthermore, masks of size $8{\times} 8$ and $32{\times} 32$ without any of the structures were generated with the proposed algorithm (see Fig.\,\ref{fig:masks_random_Jonscher_nocluster}). Using these masks improves the reconstruction quality compared to the random mask using any of the tested reconstruction algorithms.
The best average PSNR is reached using the FSR together with our structure-free mask of size $32{\times} 32$. It results in an average PSNR gain of  \SI[retain-explicit-plus]{+0.59}{\decibel} compared to a random quarter sampling mask and therefore surpasses the best $8{\times} 8$ mask from~\cite{Jonscher2014} by \SI[retain-explicit-plus]{+0.23}{\decibel}. We were able to show that our proposal that a uniform and non-regular mask is favorable is independent of the tested reconstruction algorithms.
Though higher reconstruction qualities may be achieved with a regular mask when restricting ourselves to linear interpolation, such a mask faces the severe issue that high frequencies are lost completely as a consequence of aliasing.
Future work may cover more theoretical approaches in order to find an optimal quarter sampling mask and prove its optimality.

\vspace*{-0.08cm}
\section{Acknowledgment}
\vspace*{-0.075cm}
The authors gratefully acknowledge that this work has been
supported by the Deutsche Forschungsgemeinschaft (DFG)
under contract number KA 926/5-3.

\bibliographystyle{IEEEbib}
\bibliography{literatur_jabref}

\end{document}